\documentclass[a4paper,aps]{revtex4}
\usepackage{graphicx}

\begin{document}
\title
{
NMR relaxation rate of (quasi-)one-dimensional spin gapped systems in magnetic fields
}

\author
{ 
Nobuyasu Haga and Sei-ichiro Suga
}

\affiliation
{
Deparatment of Applied Physics, Osaka University, Suita, Osaka 565-0871, Japan
}

\begin{abstract}
The NMR relaxation rates of (quasi-)one-dimensional spin-gapped systems such as the $S=1/2$ spin ladder with a diagonal interaction and the Haldane-gap system are investigated at low temperatures in magnetic fields. We show that in the regime where the gap is collapsed by magnetic fields, the NMR relaxation rate diverges with decreasing temperature and its exponent shows a behavior characteristic of the model as a function of the magnetic field.  
We compare our results with recent experimental findings for the spin ladder system ${\rm Cu_{2}(C_{5}H_{12}N_{2})_{2}Cl_{4}}$ and for the Haldane-gap system ${\rm (CH_3)_4NNi(NO_2)_3}$. 
\end{abstract}

\maketitle

Low-dimensional quantum spin systems have attracted much attention because of their novel characteristics. Intensive study has been done both theoretically and experimentally for (quasi-)one-dimensional (1D) spin-gapped systems such as the Haldane-gap system, the spin-Peierls system and the spin ladder system. 
It was shown for the 1D $S=1/2$ spin-gapped systems that the feature of each system appears not in thermodynamic quantities but in the behavior of the spin correlation functions when the energy gap is collapsed by external magnetic fields \cite{CG,ES}.
Using bosonization technique, critical and dynamical properties of the spin correlation functions for the $S=1/2$ spin-gapped systems in magnetic fields were investigated \cite{CG,FZ} and the divergence property of the NMR relaxation rate with respect to temperature $1/T_1 \sim T^{-\alpha}$ was discussed in the vicinity of the critical field $H_{c_{1}}$, where the gap is collapsed \cite{CG}. 
By combining the numerical diagonalization method with the finite-size-scaling analysis based on conformal field theory, the critical properties of several (quasi-)1D spin-gapped systems in magnetic fields were investigated between $H_{c_1}$ and the saturation field $H_{c_{2}}$, and it was shown that the systems are described by the Tomonaga-Luttinger (TL) liquid \cite{ST1,ST2,sakai,US1,US2,US3}.  Experimentally, the divergence behavior of $1/T_1$ of ${\rm Cu_{2}(C_{5}H_{12}N_{2})_{2}Cl_{4}}$, which can be described as the $S=1/2$ spin ladder system with a diagonal interaction \cite{hammar,hayward,chabou}, was observed in $H_{c_{1}} \leq H \leq H_{c_{2}}$ \cite{CB}. 
On the basis of the experimental findings, the magnetic-field dependence of the divergent exponent $\alpha$ of $1/T_1$ has been discussed quantitatively and compared with some theoretical results \cite{CG,CB,US2,GT}. 

In this Letter, we investigate the NMR relaxation rates of (quasi-)1D spin-gapped systems such as the $S=1/2$ spin ladder system with a diagonal interaction and the Haldane-gap system in magnetic fields $H_{c_{1}} \leq H \leq H_{c_{2}}$. 
Let us first consider the $S=1/2$ spin ladder system with a diagonal interaction in magnetic fields as described by the following Hamiltonian,
%
%%%%%%%%%%%%%%%%%%%%%%%%%%%%%%%%%%%%%%%%%%%%%%%%%%%%%%%%%%%%%%%%%%%%
\begin{eqnarray}
{\cal H} &=&
  J_{\bot} \sum_{i=1}^{N} \mbox{\boldmath$S$}_{i,1} \cdot 
                          \mbox{\boldmath$S$}_{i,2}  
  + J_{1} \sum_{i=1}^{N} \sum_{\mu=1}^{2} 
      \mbox{\boldmath$S$}_{i,\mu} \cdot \mbox{\boldmath$S$}_{i+1,\mu} 
  + J_{2} \sum_{i=1}^{N} \mbox{\boldmath$S$}_{i,1} \cdot 
                         \mbox{\boldmath$S$}_{i+1,2} 
  - g\mu_B H \sum_{i=1}^{N}\sum_{\mu=1}^{2}S_{i,\mu}^{z}, 
\label{eqn:ham1} 
\end{eqnarray}
%%%%%%%%%%%%%%%%%%%%%%%%%%%%%%%%%%%%%%%%%%%%%%%%%%%%%%%%%%%%%%%%%%%%
%
\noindent
where $\mu(=1, 2)$ indicates two legs of the ladder, $J_{1}$ denotes intrachain coupling, $J_{2}$ denotes the diagonal interaction and $H$ is the applied magnetic field. We set $g\mu_B=1$. 
The interchain coupling $J_{\bot}$ is assumed to be positive. 
Under the condition $J_{\bot} \gg |J_{1}|$ and $|J_{2}|$, the model can be mapped onto the 1D $S=1/2 \, XXZ$ model in effective magnetic fields \cite{Totsuka,mila,GT,FZ}:  
% 
%%%%%%%%%%%%%%%%%%%%%%%%%%%%%%%%%%%%%%%%%%%%%%%%%%%%%%%%%%%%%%%%%%%%
\begin{equation}
\tilde{\cal H}
 = \tilde{J} \sum_{i=1}^{N} \left[ 
       \tilde{S}_{i}^{x} \tilde{S}_{i+1}^{x}  
                + \tilde{S}_{i}^{y} \tilde{S}_{i+1}^{y} 
     + \Delta \tilde{S}_{i}^{z} \tilde{S}_{i+1}^{z} 
      \right] 
     - \tilde{H} \sum_{i=1}^{N} \tilde{S}_{i}^{z}, 
\label{eqn:ham2}
\end{equation}
%%%%%%%%%%%%%%%%%%%%%%%%%%%%%%%%%%%%%%%%%%%%%%%%%%%%%%%%%%%%%%%%%%%%
%
where
$\tilde{S}_i^z = 2(S_{i,1}^z-1) = 2(S_{i,2}^z-1), 
\tilde{S}_{i}^{+} = -\sqrt{2}S_{i,1}^+ = \sqrt{2}S_{i,2}^+$ and 
$\tilde{S}_{i}^{-} = -\sqrt{2}S_{i,1}^- = \sqrt{2}S_{i,2}^-$. 
The effective coupling constants and the effective magnetic field are given by using the original parameters such as 
$\tilde{J}=J_{1}-\frac{1}{2}J_{2}$, 
$\Delta \tilde{J} =\frac{1}{2}J_{1}+\frac{1}{4}J_{2}$ and 
$\tilde{H}=H-J_{\bot}-\frac{1}{2}J_{1}-\frac{1}{4}J_{2}$. 
When the anisotropy is $XY$-like, such as $-1 < \Delta \leq 1$ for $\tilde{J} > 0$, the system is gapless and is described by the TL liquid. In this case, the critical fields $H_{c_{1}}$ and $H_{c_{2}}$ of the original Hamiltonian correspond to the negative and positive saturation fields $\mp \tilde{H}_c$, respectively, where $\tilde{H}_c=|\tilde{J}|(1+\Delta)$. 
The $S=1/2$ spin ladder with strong interchain coupling ($J_{\bot}>0$) in magnetic fields can be mapped onto the $S=1/2 \, XXZ$ chain with $\Delta = 1/2$. 
The $S=1/2$ bond-alternating spin chain with strong antiferromagnetic coupling can also be mapped onto the $S=1/2 \, XXZ$ chain.
Since ${\rm Cu_{2}(C_{5}H_{12}N_{2})_{2}Cl_{4}}$ is believed to be well described by the Hamiltonian (\ref{eqn:ham1}) with parameters $J_{1}=1.0$, $J_{2}=-0.55$ and $J_{\bot}=5.5$ \cite{hayward}, the magnetic property in $H_{c_{1}} \leq H \leq H_{c_{2}}$ can be investigated by the $S=1/2 \, XXZ$ chain with the anisotropy $\Delta \sim 0.284$. 

When the NMR measurement is carried out on the nuclei located at the sites different from the electronic spins, relaxation occurs through a dipolar interaction between the nuclear and electronic spins. 
In this case, the NMR relaxation rate of the 1D $S=1/2 \, XXZ$ model takes the form 
%
%%%%%%%%%%%%%%%%%%%%%%%%%%%%%%%%%%%%%%%%%%%%%%%%%%%%%%%%%%%%%%%%%%%%
\begin{equation}
\frac{1}{T_{1}} \sim 
  T \lim_{\omega\rightarrow 0} \frac{1}{\omega} 
    \int dk\Im m\left[\chi_{\|}(k,\omega) + 
                     \chi_{\bot}(k,\omega)\right], 
\label{eqn:nmr1}
\end{equation}
%%%%%%%%%%%%%%%%%%%%%%%%%%%%%%%%%%%%%%%%%%%%%%%%%%%%%%%%%%%%%%%%%%%%
%
where $\chi_{\|}(k,\omega)$ and $\chi_{\bot}(k,\omega)$ are, respectively, the dynamical longitudinal and transverse spin susceptibilities, and are given by 
%
%%%%%%%%%%%%%%%%%%%%%%%%%%%%%%%%%%%%%%%%%%%%%%%%%%%%%%%%%%%%%%%%%%%%
\begin{equation}
%\begin{eqnarray}
\chi_{\|}(k,\omega)
= \int dx \int dt 
     \langle \tilde{S}^z(x,\tau) \tilde{S}^z(0) \rangle 
    \left| _{\tau \rightarrow it+0^+} \right. 
    {\rm e}^{-ikx} \, {\rm e}^{i \omega t}, 
\label{eqn:dsl}
%\end{eqnarray}
\end{equation}
\begin{equation}
%\begin{eqnarray}
\chi_{\bot}(k,\omega)
= \int dx \int dt 
    \langle \tilde{S}^{+}(x,\tau) \tilde{S}^{-}(0) \rangle 
    \left| _{\tau \rightarrow it+0^+} \right. 
    {\rm e}^{-ikx} \, {\rm e}^{i \omega t}. 
\label{eq:dst}
%\end{eqnarray}
\end{equation}
%
%%%%%%%%%%%%%%%%%%%%%%%%%%%%%%%%%%%%%%%%%%%%%%%%%%%%%%%%%%%%%%%%%%%%
%
In $-\tilde{H}_c \leq \tilde{H} \leq \tilde{H}_c$, the magnetization can be renormalized into one of the phase fields by bosonization treatment. Incorporating this phase field in the expressions for spin operators, we obtain the dynamical spin correlation functions as \cite{CG,GT} 
%
%%%%%%%%%%%%%%%%%%%%%%%%%%%%%%%%%%%%%%%%%%%%%%%%%%%%%%%%%%%%%%%%%%%%
%
%\begin{equation}
\begin{eqnarray}
\langle \tilde{S}^z(x,\tau) \tilde{S}^z(0) \rangle
 \sim    \tilde{m}^2 + \tilde{C}_1 r^{-2} 
       + \tilde{C}_2 r^{-\eta^z} \cos[\pi(1-2\tilde{m})x] , 
\label{eqn:rdsl}
\end{eqnarray}
%\end{equation}
%
%\begin{equation}
\begin{eqnarray}
\langle \tilde{S}^{+}(x,\tau) \tilde{S}^{-}(0) \rangle
 \sim    \tilde{D}_1 r^{-\eta^x} \cos[\pi x] 
       + \tilde{D}_2 r^{-(\eta^x + 1/\eta^x)} \cos[2\pi \tilde{m}x], 
\label{eqn:rdst}
\end{eqnarray}
%\end{equation}
%%%%%%%%%%%%%%%%%%%%%%%%%%%%%%%%%%%%%%%%%%%%%%%%%%%%%%%%%%%%%%%%%%%%
%
with $r=\sqrt{x^2+(v\tau)^2}$ and $v$ being the velocity of the elementary excitation at $T=0$. 
Note that the asymptotic decay of the dynamical spin correlation functions of the original Hamiltonian and the effective Hamiltonian shows the same behavior \cite{com}. 
By conformally mapping onto the Matsubara strip as 
$v\tau \pm ix = (v/\pi T) \sin[(\pi T/v)(v\tau \pm ix)]$, 
we obtain the dynamical spin susceptibility at finite temperature \cite{dss}. 
Paying attention to the temperature dependence of $1/T_1$, we obtain 
$1/T_1 \sim \tilde{C}_1 T + \tilde{C}_2 T^{\eta^z-1} + 
\tilde{D}_1 T^{\eta^x-1} + \tilde{D}_2 T^{\eta^x+1/\eta^x-1}$. 
Since the universal relation $\eta^x \, \eta^z = 1$ is satisfied in the TL liquid \cite{LP,Hal}, the divergence behavior appears only in the second and third terms. Thus, 
%
%%%%%%%%%%%%%%%%%%%%%%%%%%%%%%%%%%%%%%%%%%%%%%%%%%%%%%%%%%%%%%%%%%%%
\begin{equation}
\frac{1}{T_1} \sim \tilde{C}_2 T^{\eta^z-1} + \tilde{D}_1 T^{\eta^x-1},
\label{eqn:nmr2}
\end{equation}
%%%%%%%%%%%%%%%%%%%%%%%%%%%%%%%%%%%%%%%%%%%%%%%%%%%%%%%%%%%%%%%%%%%%
%
where the first term originates in the incommensurate mode of the dynamical longitudinal spin susceptibility and the last one originates in the staggered mode of the dynamical transverse spin susceptibility.

The effective Hamiltonian enables us to calculate the critical exponent 
$\eta^z$ exactly in $-\tilde{H}_c \leq \tilde{H} \leq \tilde{H}_c$ by means of the dressed charge $Z(\lambda)$  as $\eta^z = 2[Z(\Lambda)]^2$ \cite{BIK}. The dressed charge $Z(\lambda)$ is obtained from the integral equation \cite{BIK}, 
%
%%%%%%%%%%%%%%%%%%%%%%%%%%%%%%%%%%%%%%%%%%%%%%%%%%%%%%%%%%%%%%%%%%%%
%\begin{eqnarray}
$
Z(\lambda) = 1 + (1/2\pi) \int_{-\Lambda}^{\Lambda} \, d\lambda^{\prime} 
             K(\lambda-\lambda^{\prime}) Z(\lambda^{\prime}), 
$
%\label{eqn:dc}
%\end{eqnarray}
%%%%%%%%%%%%%%%%%%%%%%%%%%%%%%%%%%%%%%%%%%%%%%%%%%%%%%%%%%%%%%%%%%%%
%
where 
$K(\lambda-\lambda^{\prime}) = -\sin(4\gamma)/
  [\sinh(\lambda-\lambda^{\prime}+i2\gamma)
   \sinh(\lambda-\lambda^{\prime}-i2\gamma)]$ 
for $2\gamma \neq 0$ and 
$K(\lambda-\lambda^{\prime}) = -2/[(\lambda-\lambda^{\prime})^2+1]$ 
for $2\gamma = 0$ with $\Delta=\cos(2\gamma)$. 
The cutoff $\Lambda$ is determined by the condition for the dressed energy $\varepsilon(\pm\Lambda)=0$, where $\varepsilon(\lambda)$ is obtained from the integral equation in a given magnetic field as 
$\varepsilon(\lambda) = \varepsilon_0(\lambda) + 
  (1/2\pi) \int_{-\Lambda}^{\Lambda} \, d\lambda^{\prime} 
             K(\lambda-\lambda^{\prime}) \varepsilon(\lambda^{\prime})$ 
with 
$\varepsilon_0(\lambda) = 2\tilde{H} - 2\sin^2(2\gamma)/
  [\sinh(\lambda+i\gamma)\sinh(\lambda-i\gamma)]$ for $2\gamma \neq 0$ and 
$\varepsilon_0(\lambda) = 2\tilde{H} - 2/[\lambda^2+1/4]$ for $2\gamma = 0$. 
Note that the magnetization is obtained by the use of the dressed charge as 
$\tilde{m} = 1/2 - (1/2\pi) \int_{-\Lambda}^{\Lambda} \, d\lambda 
Z(\lambda) q_0(\lambda)$ with 
$q_0(\lambda) = \sin(2\gamma)/
  [\sinh(\lambda+i\gamma)\sinh(\lambda-i\gamma)]$ for $2\gamma \neq 0$ and 
$q_0(\lambda) = 1/[\lambda^2+1/4]$ for $2\gamma = 0$. 
In the case of $0<\Delta \leq 1$, the exponent $\eta^z$ thus obtained varies convexly downward as a function of the magnetic field and $1<\eta^z \leq 2$, while in the case of $-1<\Delta<0$, $\eta^z$ varies convexly upward and $\eta^z \geq 2$, with $\eta^z$ being $2$ at $\tilde{H}=\pm \tilde{H}_c$ in both cases. 
The behavior of $\eta^x$ can be obtained from the universal relation $\eta^x \, \eta^z = 1$. The results for $\eta^x$ and $\eta^z$ have also been confirmed by the numerical results for several systems such as the $S=1/2$ bond-alternating chain \cite{sakai}, the $S=1/2$ spin ladder \cite{US2} and the $S=1/2$ spin ladder with a diagonal interaction \cite{US2}.

On the basis of these observations, we conclude that the divergence property of the NMR relaxation rate $1/T_1$ of the system, which can be mapped onto the $S=1/2 \, XXZ$ chain with an $XY$-like anisotropy, is caused by the staggered mode of the dynamical transverse spin susceptibility,
%
%%%%%%%%%%%%%%%%%%%%%%%%%%%%%%%%%%%%%%%%%%%%%%%%%%%%%%%%%%%%%%%%%%%%
\begin{equation}
\frac{1}{T_1}  \sim  \tilde{D}_1 T^{\eta^x-1} \equiv \tilde{D}_1 T^{-\alpha}. 
\label{eqn:nmr3}
\end{equation}
%%%%%%%%%%%%%%%%%%%%%%%%%%%%%%%%%%%%%%%%%%%%%%%%%%%%%%%%%%%%%%%%%%%%
%
 The results for the exponent $\alpha$ obtained from the effective Hamiltonian and obtained numerically from the original Hamiltonian \cite{US2} are shown in Fig. 1 for the $S=1/2$ spin ladder system and the $S=1/2$ spin ladder with a diagonal interaction corresponding to ${\rm Cu_{2}(C_{5}H_{12}N_{2})_{2}Cl_{4}}$. 
The parameters for the numerical calculation of the former system are $J_{1}=1.0$, $J_{2}=0$ and $J_{\bot}=5.0$, and those for the latter system are indicated previously \cite{US2}. 
 The magnetization of the original Hamiltonian $m$ and that of the effective Hamiltonian $\tilde{m}$ satisfy the relation $2m = \tilde{m} + 1/2$. 
%
%%%%%%%%%%%%%%%%%%%%%%%%%%%%%%%%%%%%%%%%%%%%%%%%%%%%%%%%%%%%%%%
\begin{figure}[h]
\begin{center}
%\epsfigure{file=fig1.eps,height=5.5cm}
%\epsfile{file=fig1.eps,height=5.5cm}
\includegraphics[trim= 0cm 4.7cm 0cm 0cm,clip,height=5.5cm]{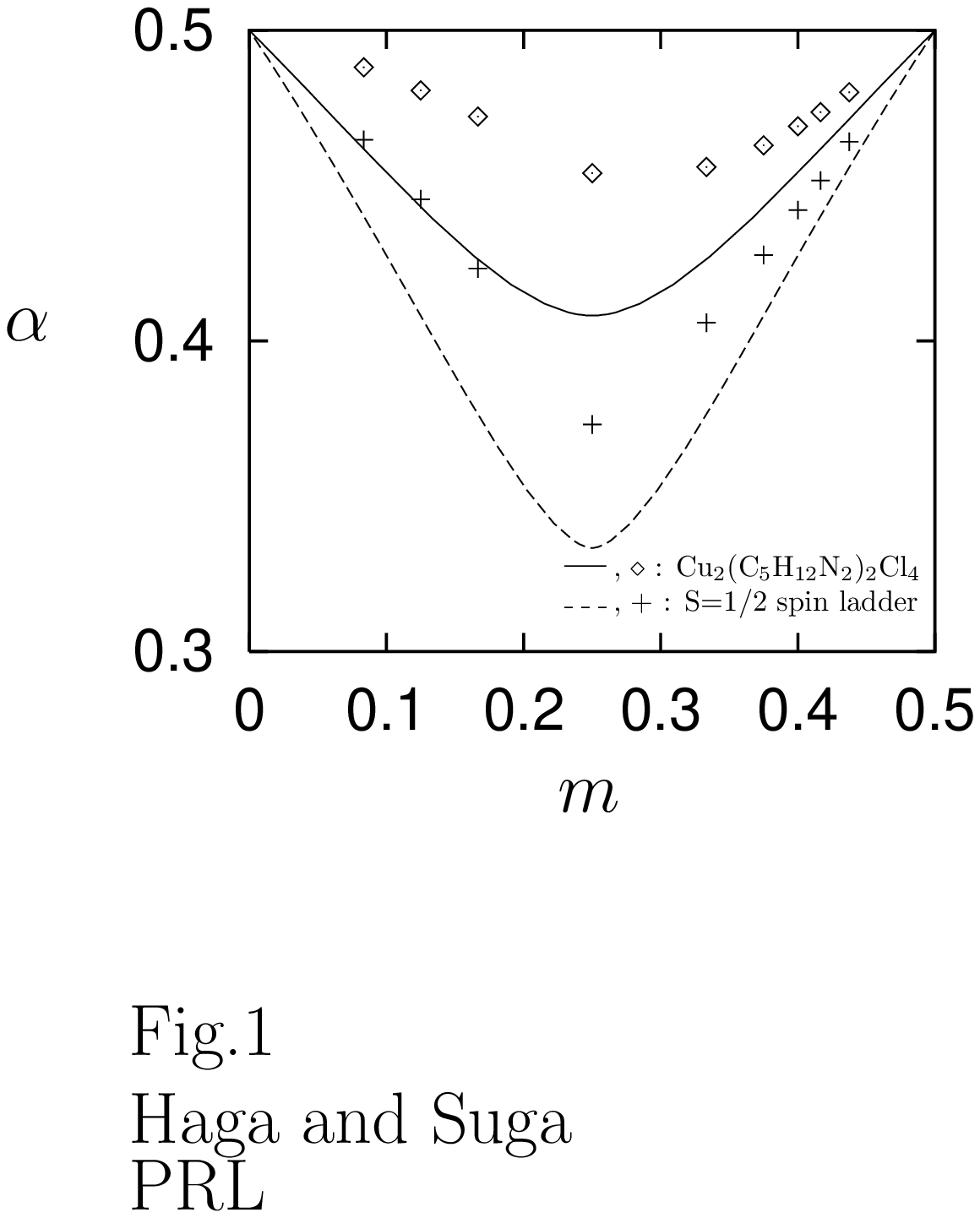}
\end{center}
\caption{Exponents of the NMR relaxation rate $1/T_1 \sim \tilde{D}_1 T^{-\alpha}$ for the $S=1/2$ spin ladder system and the $S=1/2$ spin ladder with a diagonal interaction corresponding to ${\rm Cu_{2}(C_{5}H_{12}N_{2})_{2}Cl_{4}}$ are shown as a function of magnetization $m$.  The solid line and the broken line represent the results from the 1D $S=1/2 \, XXZ$ model with $\Delta = 1/2$ and $\Delta = 0.284$, respectively. Diamonds and crosses are the results from the numerical data of ref.8. }
\label{fig1}
\end{figure}
%%%%%%%%%%%%%%%%%%%%%%%%%%%%%%%%%%%%%%%%%%%%%%%%%%%%%%%%%%%%%%%
%
The results by both methods show qualitatively the same behavior. In particular, the agreement becomes noticeable in the larger $m$ region. In the magnetic field near $H_{c_2}$, the system is nearly fully polarized and the many-body effect may be suppressed, leading to the good agreement between the results of both methods. In the magnetic field near $H_{c_1}$, on the other hand, a quantum fluctuation which relates to a gap may still survive, leading to the difference between the results by both methods.

The result for the $S=1/2$ spin ladder system is different from that obtained  based on the bosonization technique, where the exponent $\alpha=1/2$ at $m=0$ and does not change with $m$ in the vicinity of $m=0$ \cite{CG}. Furthermore, the effective TL parameter at $H=H_{c_1}$ is $\tilde{K} = 1/2$ in ref. 1, whereas $\tilde{K} = \eta^z/2 = 1$ in our result. 
At $H=H_{c_1}$ and $H_{c_2}$, the system is in the empty or fully occupied band in the fermionic language, which leads $\tilde{K} = 1$. 
Thus we conclude that in the $S=1/2$ spin ladder system, $\tilde{K}$ and $\alpha$ vary convexly downward as a function of the magnetic field with $\tilde{K} = 1$ and $\alpha=1/2$ at $H=H_{c_1}$ and $H_{c_2}$, irrespective of the strength of interchain coupling, although the magnetic field where $\tilde{K}$ and $\alpha$ take minimum values may shift in the case of weak interchain coupling.

We discuss further the magnetic-field dependence of $1/T_1$ in other systems such as the Haldane-gap system and the $S=1/2$ bond-alternating spin chain with a second-neighbor interaction. 
The latter system can be described by the Hamiltonian (\ref{eqn:ham1}). With the parameters $J_{\bot}=2.06, J_1=0.48$ and $J_2=1.94$, the Hamiltonian (\ref{eqn:ham1}) can be an effective model to describe the magnetic properties of ${\rm CuGeO_3}$ which exhibits the spin-Peierls transition \cite{SP}.

These two systems seem to be difficult to map onto the effective $S=1/2 \, XXZ$ chain. 
The critical properties of these systems in $H_{c_1} \leq H \leq H_{c_2}$ were investigated numerically, and it was revealed that the systems are described by the TL liquid \cite{ST1,ST2,US1}. 
The 1D spin systems described by the TL liquid can be mapped onto the interacting boson system, where the density operator of the boson corresponds to $\tilde{S}^z$ of the effective 1D $S=1/2 \, XXZ$ model and the creation and annihilation operators of the boson correspond to $\tilde{S}^{\pm}$. 
The dynamical correlation functions of these boson operators take the same asymptotic forms as (\ref{eqn:rdsl}) and (\ref{eqn:rdst}).   
Therefore, the result for $1/T_1$ shown in (\ref{eqn:nmr2}), which has been derived based on (\ref{eqn:rdsl}) and (\ref{eqn:rdst}), holds generally in 1D spin systems described by the TL liquid.   
Then, we investigate the divergence behavior of $1/T_1$ in 1D spin systems described by the TL liquid, which cannot be mapped onto the effective 1D $S=1/2 \, XXZ$ model, using the result (\ref{eqn:nmr2}). 

We first consider the Haldane-gap system. 
 The numerical results indicate that $\eta^x \leq 1/2$ and $\eta^z \geq 2$ with $\eta^x = 1/2$ and $\eta^z = 2$ at $H_{c_1}$ and $H_{c_2}$ \cite{ST1,ST2}. 
Judging from these results and (\ref{eqn:nmr2}), we conclude that the divergence behavior of $1/T_1$ is caused by the staggered mode of the dynamical transverse spin susceptibility and $1/T_1$ is expressed as shown in (\ref{eqn:nmr3}) \cite{SA}. The results for $\alpha$ in the Haldane-gap system are shown in Fig. 2 using the numerical results in ref. 4. 
%
%%%%%%%%%%%%%%%%%%%%%%%%%%%%%%%%%%%%%%%%%%%%%%%%%%%%%%%%%%%%%%%
\begin{figure}[h]
\begin{center}
\includegraphics[trim= 0cm 5.8cm 0cm 0cm,clip,height=5.5cm]{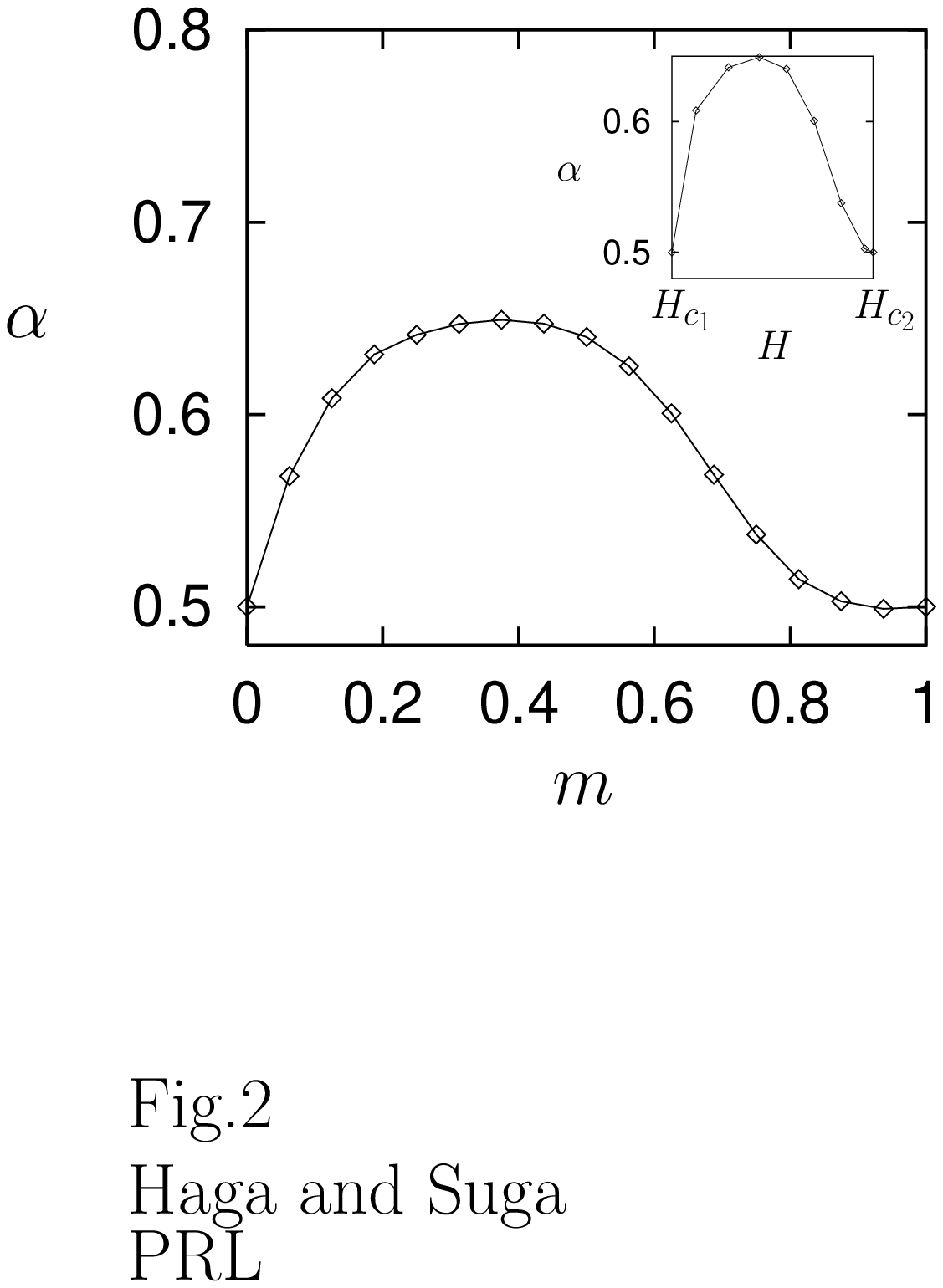}
\end{center}
\caption{\small Exponent of $1/T_1 \sim \tilde{D}_1 T^{-\alpha}$ for the Haldane-gap system is shown as a function of $m$. The inset shows the $H$ dependence of $\alpha$.}
%The numerical data of ref.4 are used. }
\label{fig2}
\end{figure}
%%%%%%%%%%%%%%%%%%%%%%%%%%%%%%%%%%%%%%%%%%%%%%%%%%%%%%%%%%%%%%%
%
 The exponent $\alpha$ increases from $1/2$,  and takes a maximum of $\sim 0.65$ around $m \sim 0.4$. At $H=H_{c_2}$, $\alpha=1/2$. 
 Quite recently, the divergence behavior of $1/T_1$ in the Haldane-gap system ${\rm (CH_3)_4NNi(NO_2)_3}$ was observed experimentally at low temperatures \cite{Goto}. The exponent of $1/T_1$ with respect to temperature has been estimated as 0.5 at $m \sim 0$ and as 0.7 at $m \sim 0.2$. These experimental results agree with the results shown in Fig. 2 quantitatively.

We next consider the $S=1/2$ bond-alternating spin chain with a second-neighbor interaction with the parameters $J_{\bot}=2.06, J_1=0.48$ and $J_2=1.94$. 
The critical exponents $\eta^x$ and $\eta^z$ show that $\eta^x \geq 1$ and $\eta^z \leq 1$ in $0.1<m \leq m_c$, and $1/2 \leq \eta^x \leq 1$ and $1 \leq \eta^z \leq 2$ in $m_c \leq m \leq 1/2$, where $m_c \sim 0.35$ and $\eta^x=\eta^z=1$ at $m=m_c$ \cite{US1}. 
From (\ref{eqn:nmr2}), we conclude that 
$1/T_1 \sim \tilde{C}_2 T^{-(1-\eta^z)}$ in $0.1<m \leq m_c$ and 
$1/T_1 \sim \tilde{D}_1 T^{-(1-\eta^x)}$ in $m_c \leq m \leq 1/2$. 
 The results for the divergent exponent $\alpha$ in this system are shown in Fig. 3 using the numerical results in ref. 7. 
%
%%%%%%%%%%%%%%%%%%%%%%%%%%%%%%%%%%%%%%%%%%%%%%%%%%%%%%%%%%%%%%%
\begin{figure}[h]
\begin{center}
\includegraphics[trim= 0cm 5.9cm 0cm 0cm,clip,height=5.5cm]{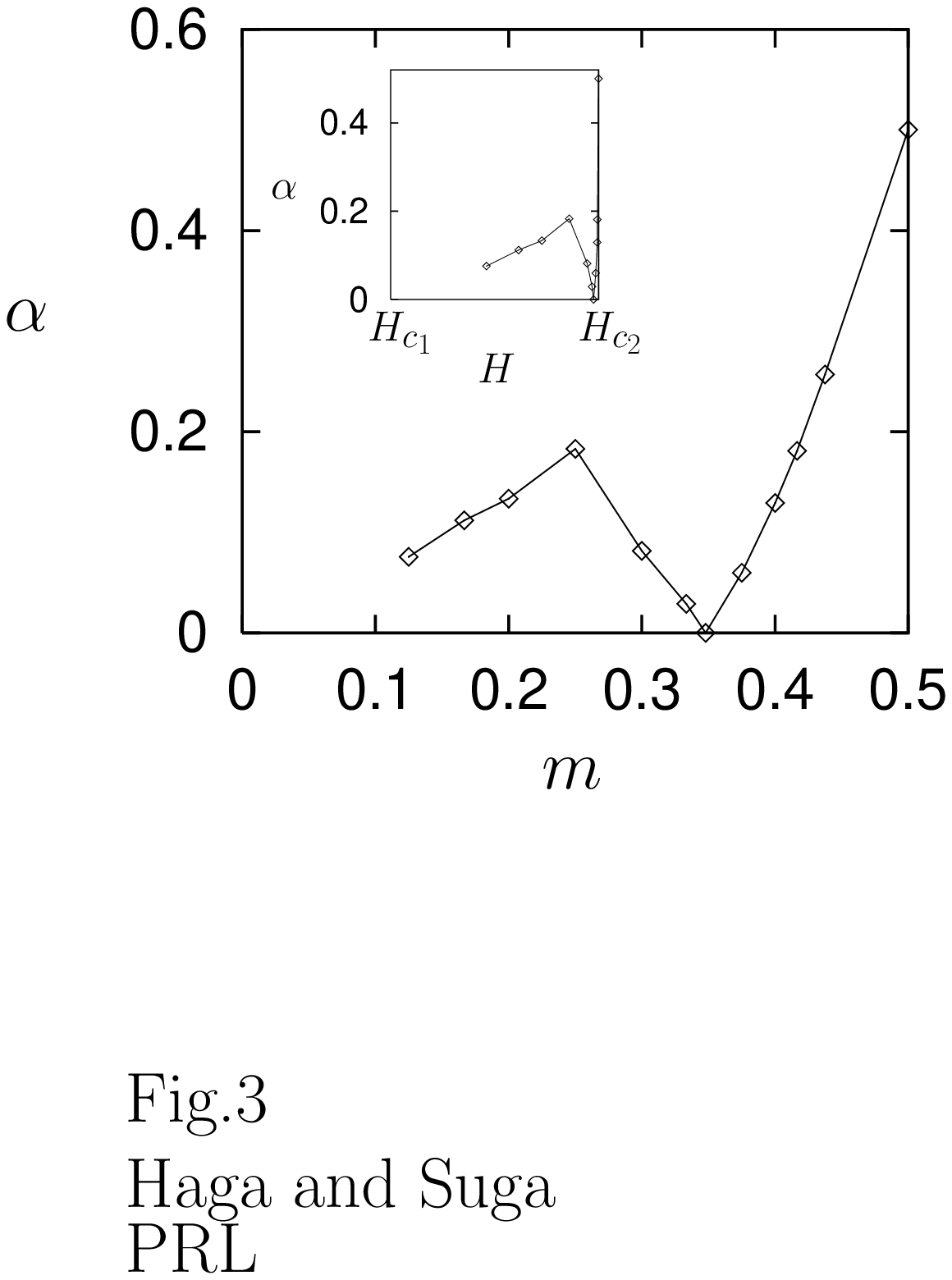}
\end{center}
\caption{\small Exponent of $1/T_1$ for the $S=1/2$ bond-alternating spin chain with a second-neighbor interaction is shown as a function of $m$. The parameters are $J_{\bot}=2.06, J_1=0.48$ and $J_2=1.94$. The inset shows the $H$ dependence of the exponent. }
%The numerical data of ref.7 are used. }
\label{fig3}
\end{figure}
%%%%%%%%%%%%%%%%%%%%%%%%%%%%%%%%%%%%%%%%%%%%%%%%%%%%%%%%%%%%%%%
%
The results indicate that the leading contribution of the spin correlation function is the incommensurate mode of $\langle \tilde{S}^z(x,\tau) \tilde{S}^z(0) \rangle$ in $0.1<m \leq m_c$, and is the staggered mode of $\langle \tilde{S}^{+}(x,\tau) \tilde{S}^{-}(0) \rangle$ in $m_c \leq m \leq 1/2$. 
At $H = H_{c_1}$, the system is in the empty or fully occupied band and $\tilde{K} = \eta^z/2 = 1$. Therefore, it is expected that near $H = H_{c_1}$, the leading contribution is the staggered mode of $\langle \tilde{S}^{+}(x,\tau) \tilde{S}^{-}(0) \rangle$ and that the dominant mode changes again at the magnetic field between $H_{c_1}$ and $0.1$. 
In the region where the longitudinal incommensurate mode is dominant, the system has a tendency to form an incommensurate spin order with period $1/m$. 
This tendency may relate to the property of the incommensurate phase of ${\rm CuGeO_3}$ in magnetic fields \cite{sl}.

The behavior of $\eta^x$ and $\eta^z$ in $0.1<m \leq m_c$ is most likely brought about by the long-range interaction: the second-neighbor antiferromagnetic interaction. Hence, in systems such as the $S=1/2$ $n$-leg spin ladder system with diagonal interactions, the incommensurate mode of the longitudinal spin correlation function can be dominant in an intermediate range between $H_{c_1}$ and $H_{c_2}$, while the staggered mode of the transverse one can be dominant near $H_{c_1}$ and $H_{c_2}$. This characteristic behavior is detectable by the NMR relaxation rate in magnetic fields, if the probed nucleus is at a site different from electronic spins.

Finally, we comment on the NMR relaxation rate of the Heisenberg alternating-spin chain with $S=1$ and $S=1/2$ in magnetic fields. It was shown numerically that the system belongs to the universality class of the TL liquid when $1/4 < m < 3/4$ \cite{al}. Using the numerical results for $\eta^x$ and $\eta^z$ \cite{al} and (\ref{eqn:nmr2}), we see that the divergence behavior of $1/T_1$ is dominated by the staggered mode of the dynamical transverse spin susceptibility, and the exponent seems to take a minimum of $\sim 0.4$ around $m = 0.4$. 

In summary, we have systematically investigated the NMR relaxation rates of (quasi-)1D spin-gapped systems at low temperatures in $H_{c_1} \leq H \leq H_{c_2}$. The results are summarized in Table I. The divergent exponent of $1/T_1$ with respect to temperature shows a feature characteristic of the model, reflecting the nature of the TL liquid.  

We would like to thank H. Frahm, T. Goto, N. Kawakami and A. Tomiyama for their useful comments and valuable discussion. 
This work was supported by Grant-in-Aid no. 10640344 for Scientific Research from the Ministry of Education, Science, Sports and Culture, Japan.

%
%%%%%%%%%%%%%%%%%%%%%%%%%%%%%%%%%%%%%%%%%%%%%%%%%%%%%%%%%%%%%%%
\begin{table}[h]
\caption{The divergent exponents of $1/T_1$ in magnetic fields are summarized.  AF (F) Ladder denotes the $S=1/2$ ladder with antiferromagnetic (ferromagnetic) interchain couplings, AF-AF(-F) BA denotes the $S=1/2$ antiferromagnetic-antiferromagnetic(-ferromagnetic) bond-alternating chain and Frustrated BA denotes the system with $J_{\bot}=2.06, J_1=0.48$ and $J_2=1.94$.}
\begin{center}
\begin{tabular}{cccc} \hline \hline
Model & \makebox[2em]{}at $H_{c_1}$ and $H_{c_2}$ & \makebox[2em]{}Dominant mode & \makebox[2em]{}$H_{c_1}<H<H_{c_2}$ \\ \hline
${\rm Cu_{2}(C_{5}H_{12}N_{2})_{2}Cl_{4}}$ & \makebox[2em]{}1/2 & \makebox[2em]{}transverse staggered mode & \makebox[2em]{}convex downward \\
AF Ladder & \makebox[2em]{}1/2 & \makebox[2em]{}transverse staggered mode & \makebox[2em]{}convex downward \\
AF-AF BA & \makebox[2em]{}1/2 & \makebox[2em]{}transverse staggered mode & \makebox[2em]{}convex downward \\
Alternating-spin chain & \makebox[2em]{}1/2 & \makebox[2em]{}transverse staggered mode & \makebox[2em]{}convex downward \\
Haldane Gap & \makebox[2em]{}1/2 & \makebox[2em]{}transverse staggered mode & \makebox[2em]{}convex upward \\
AF-F BA & \makebox[2em]{}1/2 & \makebox[2em]{}transverse staggered mode & \makebox[2em]{}convex upward \\
F Ladder & \makebox[2em]{}1/2 & \makebox[2em]{}transverse staggered mode & \makebox[2em]{}convex upward \\
Frustrated  BA & \makebox[2em]{}1/2 & \makebox[2em]{}dominant mode changes & \makebox[2em]{}depending on the mode \\
\hline \hline
\end{tabular}
\end{center}
\end{table}
%%%%%%%%%%%%%%%%%%%%%%%%%%%%%%%%%%%%%%%%%%%%%%%%%%%%%%%%%%%%%%%
%

%%%%%%%%%%%%%%%%%%%%%%%%%%%%%%%%%%%%%%%%%%%%%%%%%%%%%%%%%%%%%%%%%%%%%
%                        REFERENCES                                 %
%%%%%%%%%%%%%%%%%%%%%%%%%%%%%%%%%%%%%%%%%%%%%%%%%%%%%%%%%%%%%%%%%%%%%
%
%\begin{references}

%\bibitem{tag} Fake bibitem.
%\end{references}
%
%
%%%%%%%%%%%%%%%%%%%%%%%%%%%%%%%%%%%%%%%%%%%%%%%%%%%%%%%%%%%%%%%%%%%%%


\begin{thebibliography}{99}
\bibitem{CG}
R. Chitra and T. Giamarchi: Phys. Rev. B {\bf 55} (1997) 5816.
\bibitem{ES}
N. Elstner and R. R. P. Singh: Phys. Rev. B {\bf 58} (1998) 11484.
\bibitem{FZ}
A. Furusaki and S.-C. Zhang: Phys. Rev. B {\bf 60} (1999) 1175.
\bibitem{ST1}
T. Sakai and M. Takahashi: J. Phys. Soc. Jpn. {\bf 60} (1991) 3615.
\bibitem{ST2}
T. Sakai and M. Takahashi: Phys. Rev. B {\bf 43} (1991) 13383.
\bibitem{sakai}
T. Sakai: J. Phys. Soc. Jpn. {\bf 64} (1995) 251.
\bibitem{US1}
M. Usami and S. Suga: Phys. Lett. A {\bf 240} (1998) 85.
\bibitem{US2}
M. Usami and S. Suga: Phys. Rev. B {\bf 58} (1998) 14401.
\bibitem{US3}
M. Usami and S. Suga: Phys. Lett. A {\bf 259} (1999) 53. 
\bibitem{hammar}
P. R. Hammar and D. H. Reich: J. Appl. Phys. {\bf 79} (1996) 5392.
\bibitem{hayward}
C. A. Hayward, D. Poilblanc and L. P. L\'evy: Phys. Rev. B {\bf 54} (1996) 12649.
\bibitem{chabou}
G. Chaboussant, P. A. Crowell, L. P. L\'evy, O. Piovesana, A. Madouri and D. Mailly: Phys. Rev. B {\bf 55} (1997) 3049.
\bibitem{CB}
G. Chaboussant, Y. Fagot-Revrat, M.-H. Julien, M. E. Hanson, C. Berthier, M. Horvati\'{c}, L. P. L\'{e}vy and O. Piovesana: Phys. Rev. Lett. {\bf 80} (1998) 2713.
\bibitem{GT}
T. Giamarchi and A. M. Tsvelik: Phys. Rev. B {\bf 59} (1999) 11398.
\bibitem{Totsuka}
K. Totsuka: Phys. Rev. B {\bf 57} (1998) 3454.
\bibitem{mila}
F. Mila: Eur. Phys. J. B {\bf 6} (1998) 201.
\bibitem{com}
The dynamical spin correlation functions of the original Hamiltonian are given by 
$
\langle \phi^z(x, \tau) \phi^z(0) \rangle 
= C_0 m^2 + C_1 r^{-2} + C_2 r^{-\eta^z} \! \cos[2\pi m x] 
$
and 
$
\langle \phi^+(x, \tau) \phi^-(0) \rangle 
= D_1 r^{-\eta^x} \! \cos[\pi x] + D_2 r^{-(\eta^x + 1/\eta^x)} \! \cos[\pi(1-2m)x], 
$
where a spin operator $\mbox{\boldmath$\phi$}_i$ is composed of two $S=1/2$ operators $\mbox{\boldmath$S$}_{i,1}$ and $\mbox{\boldmath$S$}_{i,2}$. 
\bibitem{dss}
The analytical results for the dynamical spin susceptibility of the antiferromagnetic Heisenberg chain at finite temperature were investigated in H. J. Schulz: Phys. Rev. B {\bf 34} (1984) 6372; S. Sachdev, T. Senthil and R. Shankar: Phys. Rev. B {\bf 50} (1994) 258.
Using these results, the NMR relaxation rate for half-integer Heisenberg spin chains in the absence of magnetic fields was calculated in S. Sachdev: Phys. Rev. B {\bf 50} (1994) 13006.
\bibitem{LP}
A. Luther and I. Peschel: Phys. Rev. B {\bf 12} (1975) 3908.
\bibitem{Hal}
F. D. M. Haldane: Phys. Rev. Lett. {\bf 45} (1980) 1358; Phys. Lett. A {\bf 81} (1981) 153; J. Phys. C {\bf 14} (1981) 2585. 
\bibitem{BIK}
N. M. Bogoliubov, A. G. Izergin and V. E. Korepin: Nucl. Phys. B {\bf 275} (1986) 687. 
\bibitem{SP}
G. Castilla, S. Chakravarty and V. J. Emery: Phys. Rev. Lett. {\bf 75} (1995) 1823.
\bibitem{SA}
On the basis of the nonlinear $\sigma$ model, the temperature dependence of the NMR relaxation rate of the Haldane-gap system in $H > H_{c_1}$ was derived as $1/T_1 \sim T^{\eta^x-1}$ in 
J. S. Sagi: Ph. D. Thesis, the University of British Columbia (1995), cond-mat/9512161. 
\bibitem{Goto}
T. Goto, Y. Fujii, Y. Shimaoka, T. Maekawa and J. Arai: to be published in Physica B (2000); T. Goto (private communication). 
\bibitem{sl}
See, for example, G. S. Uhrig: in {\it Advances in Solid State Physics}, ed. B. Kramer, Vol. 39 (Vieweg Verlag, Braunschweig, 1999), p. 291, cond-mat/9902272, and references therein.
\bibitem{al}
T. Kuramoto: J. Phys. Soc. Jpn. {\bf 67} (1998) 1762.
\end{thebibliography}
\end{document}